# POSTER: Self-Healing Mechanisms for Software-Defined Networks

José Sánchez[1,2], Imen Grida Ben Yahia[1], Noël Crespi[2]
[1] Orange Labs, Paris, France
[2] Telecom SudParis, Evry, France

*Abstract*— Operators perceive programmable networks brought by Software Defined Networks (SDN) as cornerstone to decrease the time to deploy new services, to augment the flexibility and to adapt network resources to customer needs at runtime. However, despite the vulnerabilities identified due that the intelligence is centralized on SDN, its research is more centered on forwarding traffic and reconfiguration issues, not considering to a great extent the fault management aspects of the control plane. This paper provides SDN with fault management capabilities by using autonomic principles like self-healing mechanisms. We propose a generic self-healing approach that relies on a Bayesian Networks for the diagnosis block and we applied this algorithm to a centralized SDN infrastructure to prove its functioning in the presence of faults.

## I. MOTIVATION AND APPROACH

Today's networks are complex sets of heterogeneous and vendor-dependent equipment that use proprietary management applications. We characterize this complexity by the high rate of events occurred, the enormous amount of involved equipment, the short life-cycle and the continuously growing introduction of new technologies and services. Such diversity and complexity implies a huge cost and effort to deploy new services and manage all the equipment, what prevents Operators from improving network and service quality and reliability to satisfy the market needs and cope with the aggressive competitiveness of the industry. One symptom of this complexity is the overwhelming and unmanageable amount of alarms received by medium-sized operational teams, estimated in millions per day [1]. This urgently requires automation and intelligence to keep this management under control.

The arrival of new technologies such as Software Defined Networks (SDN), Network virtualization, Networks Function Virtualization (NFV)[2] and the Cloud, will suppose a shift from configurable to programmable networks, which is an enabler to introduce services and cost reduction, but, on the other hand, it emphasizes the need for autonomics.

SDN paves the way towards programmable networks, which would let change the network behavior by reprogramming the equipment in a flexible way as response to dynamic changes.

SDN has diverse definitions [2][3][15], but all them propose a clear separation of the control plane from the data plane. We characterize SDN by:

-A programmable control of the network through open Application Programming Interfaces (APIs).

-The centralization of the intelligence in the control plane to offer fine-grained control of the network.

-The Service Abstraction Layer (SAL) hides the details of lower SDN layers.

However, this centralization of the intelligence on SDN jeopardizes the control plane, what has disastrous consequences on the data plane. Thus, the resiliency of the controller is paramount, because it orchestrates the data plane by installing the forwarding flows. Despite this importance, no so much research has been found that surveyed fault management for SDN. Nevertheless, exists some related work [5-12] to offer SDN with fault management capabilities, but counts on several limitations:
(i) this work is OpenFlow oriented, so the recovery is an OpenFlow dependent mechanism that reconfigures or provides alternative paths to the affected OpenFlow switches.
(ii) this work does not consider malfunctions in the control plane to a great extent.
(iii) lack of an integrated fault management for legacy equipment and OpenFlow-based.
(iv) most of the recovery mechanisms are for in-band SDN architectures.

The introduction of autonomic principles on SDN guarantees an immediate detection and reparation of any malfunction. Autonomics [13] enables advanced automation and intelligence on current fault management approaches to perform them more efficiently and better. The autonomic properties are known as self-x, including self-healing. We define autonomic fault management based on self-healing systems. Self-healing systems recover the network from abnormal states and bring it back to a normal state.

A self-healing system is a closed-loop system composed of detection, diagnosis and recovery blocks. The Self-Healing mechanism receives the network alarms through its sensors to detect faults and runs this closed-loop to diagnose the root cause and applies the proper recovery strategy through the actuators. We consider that self-healing principles are mature enough and supported by many researches [14] that place them as the proper solution for enabling automation and intelligence on SDN architectures. We center our main research question on bringing these autonomic principles to ensure the resiliency and robustness of centralized SDN out-of-band architectures, where, besides OpenFlow switches, also other legacy equipment would take part of the data plane (e.g. routers or APs).

## II. ENVISAGED SELF-HEALING FRAMEWORK FOR SDN

We envision a generic self-healing module for SDN architectures (Fig. 1), capable of monitoring the services deployed over the control plane, the SDN controller, and its underlying network. Although we can place the self-healing module at the points (1), (2), (3) or (4), we initially consider the (2), due to the fact that the self-healing module requires a continuous interaction with the controller to retrieve controller performance, network performance and the topology. The self-healing module also interacts with the plane (1) through the Service Manager (SM) to receive information about the state of the services.

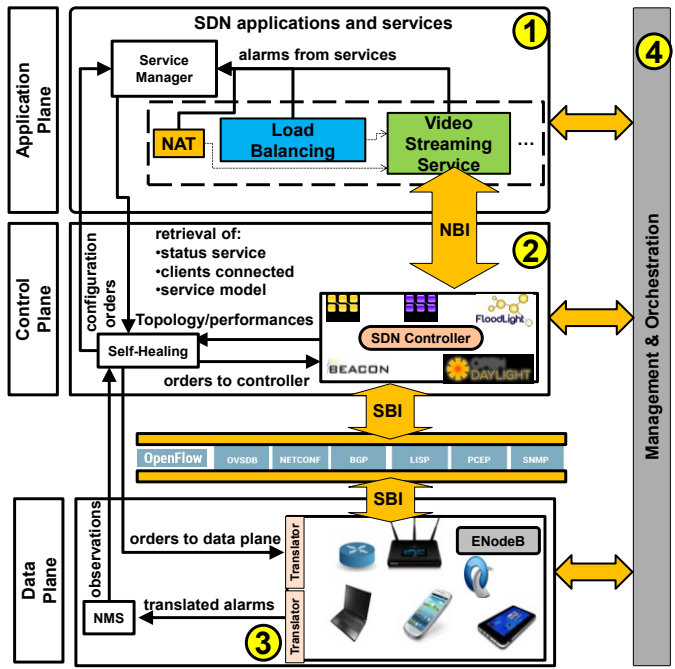

Fig. 1. Self-Healing proposed module for SDN

The Network Management System (NMS) and the SM ensure the detection block. The former takes the alarms from the network elements and the latter takes information from the services deployed. This information has the status of the services, the list of clients connected to each service as well as the service model. We consider three different levels of alarms:
-Service-related alarms: concern malfunctions on the services of the application plane such as misconfigurations, errors or SLA violations.
-Transport-related alarms: concern malfunctions on the forwarding mechanisms of the control plane such as misconfiguration, errors or QoS violations like dropped packets.
-Physical-related alarms: concern physical failures on the equipment.

The diagnosis block uses on a Bayesian Networks algorithm that incorporates the observations coming from the translated alarms from the NMS and SM and adds them as evidences for diagnosing the root cause. This block uses the network topology information, given by the SDN controller, for building the Bayesian Network graph. Also, it receives the status of the services, given by the SM. Bayesian Networks is a model-based algorithm that models complex dependencies of the network under uncertainty. Many researches on network diagnosis support this algorithm [16] and it counts on many free-licensed toolboxes [17][18][19].

The recovery block takes the diagnosis result coming from the diagnosis block and launches the appropriate recovery strategy that launches reconfiguration orders to the plane causing the malfunction (the controller, the SM, and the data plane) to re-establish the affected service.

As a first part of our work, we integrated our self-healing module in an in-band centralized SDN architecture, whose by using POX SDN controller POX [20], under a video streaming service delivered through a fixed network topology emulated on Mininet [21]. The self-healing module can detect, diagnose and repair faults of different nature (e.g. physical failures, streaming services, OpenFlow crashes and traffic drops on any interface). If the streaming service behaves abnormally, the self-healing detection block detects this behaviour, diagnoses the root cause and suggests the corresponding actions to re-establish the streaming service.

## III. FUTURE WORK

Catering for resiliency capabilities to SDN is a novel topic, but more concretely, this approach fills the gap between autonomics, self-healing in particular, and SDN. We include a Bayesian Network-based algorithm on SDN, to empower this architecture to detect disruptions on the application plane, the control plane and the data plane at run-time.

We aim to extend our developed framework to monitor virtualized services like OpenIMSCore [22], OpenLTE [23] or OpenVoLTE [24] deployed over an SDN OpenDaylight [25] controller. We identifiy the following research lines:

-Learning of topology: Extraction of the topology from the SDN controller and building of the dependency graph automatically from the topology.

- Accuracy on diagnosis: Reduction of the uncertainty of the diagnosis through a closed-loop that considers the effects of the proposed actions.
- Proactive framework: Inclusion of QoS measurements on both control and application planes to prevent future malfunctions by monitoring degradations.
- Recovery techniques: Inclusion of different techniques such as reconfiguration orders, swapping mechanisms for the controller, alternative forwarding for OpenFlow-enabled switches or load balancing on access points.
- Semantic translation for alarms: Implementation of translation mechanisms for the alarms emitted from the data plane that depend on different southbound protocols.